\documentclass[12pt,fleqn]{article}
\usepackage{amsfonts,amssymb,cite}
\usepackage{epsf,graphicx}

\newcommand{\bls}[1]{\renewcommand{\baselinestretch}{#1}}

\def\noi{\noindent}

\newcommand{\Title}[1]{\noi {{\Large\bf #1}}\\[1ex]}

\newcommand{\Author}[2]{\noi{\large\bf #1}\\[2ex]\noi{\small\it #2}\\}

\newcommand{\Abstract}[1]{\vskip 2mm \begin{center}
        \parbox{16.4cm}{\small\noi #1} \end{center}\medskip}

\newcommand{\foom}[1]{\protect\footnotemark[#1]}

\def\email#1#2{\footnotetext[#1]{e-mail: #2}\addtocounter{footnote}{1}}

\def\nqq{\hspace*{-2em}}

\def\cm{\hspace*{1cm}}

\def\Jl#1#2{#1 {\bf #2},\ }

\def\ApJ#1 {\Jl{Astroph. J.}{#1}}
\def\CQG#1 {\Jl{Class. Quantum Grav.}{#1}}
\def\DAN#1 {\Jl{Dokl. AN SSSR}{#1}}
\def\GC#1 {\Jl{Grav. Cosmol.}{#1}}
\def\GRG#1 {\Jl{Gen. Rel. Grav.}{#1}}
\def\JETF#1 {\Jl{Zh. Eksp. Teor. Fiz.}{#1}}
\def\JETP#1 {\Jl{Sov. Phys. JETP}{#1}}
\def\JHEP#1 {\Jl{JHEP}{#1}}
\def\JMP#1 {\Jl{J. Math. Phys.}{#1}}
\def\NPB#1 {\Jl{Nucl. Phys. B}{#1}}
\def\NP#1 {\Jl{Nucl. Phys.}{#1}}
\def\PLA#1 {\Jl{Phys. Lett. A}{#1}}
\def\PLB#1 {\Jl{Phys. Lett. B}{#1}}
\def\PRD#1 {\Jl{Phys. Rev. D}{#1}}
\def\PRL#1 {\Jl{Phys. Rev. Lett.}{#1}}

\def\lal{&&\nqq {}}
\def\eq{Eq.\,}
\def\eqs{Eqs.\,}
\def\beq{\begin{equation}}
\def\eeq{\end{equation}}
\def\bear{\begin{eqnarray}}
\def\bearr{\begin{eqnarray} \lal}
\def\ear{\end{eqnarray}}
\def\earn{\nonumber \end{eqnarray}}

\def\yyy{\\[5pt] \lal }

\def\dst{\displaystyle}

\def\fracd#1#2{{\dst\frac{#1}{#2}}}

\def\Half{{\fracd{1}{2}}}

\def\e{{\,\rm e}}
\def\d{\partial}

\def\tr{\mathop{\rm tr}\nolimits}

\def\const{{\rm const}}

\def\mN{_{\mu}^{\nu}}

\def\vac{{}_{\rm (vac)}}

\def\kappa{\varkappa}

\def\N{{\mathbb N}}

\def\sph{spherically symmetric}
\def\ssph{static, spherically symmetric}
\def\bh{black hole}
\def\bhs{black holes}

\bls {1.1}
\begin{document}

\Title{General static black holes in matter}

\Author{K.A. Bronnikov\foom 1}
{Center for Gravitation and Fundamental Metrology, VNIIMS, 46
  Ozyornaya St., Moscow 119361, Russia;\\ Institute of Gravitation and
  Cosmology, PFUR, 6 Miklukho-Maklaya St., Moscow 117198, Russia}
\email 1 {kb20@yandex.ru}

\Author{Oleg B. Zaslavskii\foom 2}
{Astronomical Institute of Kharkov V.N. Karazin National
  University, 35 Sumskaya St., Kharkov, 61022, Ukraine\,\,}
\email 2 {ozaslav@kharkov.ua}

\Abstract
{For arbitrary static space-times, it is shown that an equilibrium between a
  Killing horizon and matter is only possible for some discrete values of
  the parameter $w = p_1/\rho$, where $\rho$ is the density and $p_1$ is
  pressure in the direction normal to the horizon. In the generic situation
  of a simple (non-extremal) horizon and the slowest possible density
  decrease near the horizon, this corresponds to $w = -1/3$, the value
  known for a gas of disordered cosmic strings. An admixture of ``vacuum
  matter'', characterized by $w=-1$ and nonzero density at the horizon, is
  also admitted. This extends the results obtained previously for \ssph\
  space-times. A new feature as compared to spherical symmetry is that
  higher-order horizons can exist in the absence of vacuum matter if the
  horizon is a surface of zero curvature, which can occur, e.g., in
  cylindrically symmetric space-times.  }


\section{Introduction}

  It is generally believed that, in the course of gravitational collapse,
  as a \bh\ is being formed, nonspherical perturbatons die out. However,
  under real astrophysical conditions, \bhs\ are surrounded by different
  kinds of celestial bodies and distributed matter (stars in double and
  multiple stellar systems, gas clouds etc.) Even if a \bh\ can be regarded
  static in a good approximation, these bodies distort its spherical shape,
  and not always such deflections are small. On the other hand, distributed
  matter can either be in equilibrium with the \bh\ or fall on it. Conditions
  of such equilibrium are of substantial interest.

  It is also worth noting that the generic properties of Killing horizons (not
  necessarily \sph) play an important role in the (not yet well understood)
  relationship between the horizon symmetry and black hole entropy (see,
  e.g., \cite{viss} and references therein). A discussion of \bh\ entropy in
  connection with statistical properties of the surrounding matter also
  requires a knowledge of which kinds of matter can be in equilibrium with
  the horizon.

  In our previous paper \cite{bz2}, we studied the equilibrium conditions
  between macroscopic matter and \ssph\ \bhs. Using near-horizon expansions
  in the Einstein equations and the conservation law for matter, we found
  that only some discrete values of the parameter $w = p_r/\rho$ (the radial
  pressure to density ratio) are compatible with such equilibrium, and its
  generic values are $w=-1$ (which is well known \cite{FroNov, viss}) and $w
  = -1/3$ (which corresponds to a gas of cosmic strings if this matter is
  isotropic); normal ($w \geq 0$) or phantom ($w < -1$) matter is excluded.

  In the present paper we extend this approach to general static \bhs. It
  turns out that almost all conclusions, made previously for the case of
  spherical symmetry, remain valid when this requirement is relaxed.

\section{Basic equations}

  We consider the generic static metric which can be written, in terms of
  the so-called Gaussian coordinates, in the form\footnote
  	{We use the units $c= \hbar = G =1$.}
\beq
	ds^2 = N^2 dt^2 - dl^2 - \gamma _{ab}dx^{a}dx^{b},\cm a,b=2,3,
\eeq
  where $N$ and elements of the 2-metric $\gamma_{ab}$ are functions of
  $l, x^2, x^3$. Horizons, if any, are located at surfaces where $N=0$.

  Let us assume that the stress-energy tensor (SET) of matter has the
  components $T_0^0 = \rho$, $T_1^1 = p_1$ ($x^1 = l)$, $T_0^i = 0$
  $(i=1,2,3)$. Then the $l$ component of the conservation law $\nabla_\nu
  T\mN =0$ gives
\beq                    				\label{cons}
	p'_1 + (p_{1}+\rho )\frac{N'}{N}
	+ \biggl(p_1 \frac{\gamma'}{2\gamma} + K_{ab} T^{ab}\biggr)=0,
\eeq
  where the prime denotes $\d/\d l$, $\gamma = \det (\gamma_{ab})$, and
\beq
   	K_{ab} = -\frac{1}{2} \frac{\d \gamma_{ab}}{\d l}         \label{K}
\eeq
  is the extrinsic curvature tensor of the surfaces $t = \const$,
  $l = \const$ showing how they are embedded in the outer three-space.

  Using the explicit expressions for the Einstein tensor given in \cite{viss},
  we can write two components of the Einstein equations as follows:
\bearr                                                          \label{1-0}
	G_1^1 - G_0^0 = \frac{\Delta _{2}N}{N} + K'-\frac{N'}{N}
			K - K_{ab}K^{ab} = 8\pi (p_1 + \rho ),
\yyy                                                             \label{11}
	G_1^1 = -\frac{1}{2}R_{\parallel} - \Half K_{ab}K^{ab}
	+ \Half K^2 + \frac{\Delta_2 N}{N} - \frac{KN'}{N} = 8\pi p_1,
\ear
  where $K = \tr K = \gamma^{ab} K_{ab}$, $R_{\parallel}$ is the scalar
  curvature of the two-surface $t=\const$, $l = \const$, and $\Delta_2$
  is the two-dimensional Laplacian corresponding to the metric $\gamma_{ab}$
  on such a surface.

\section {Horizons and their neighborhood}

  Considering a two-dimensional $(t, l)$ section of our manifold at fixed
  values of the coordinates $x^a$, we can introduce the so-called quasiglobal
  coordinate $u$ by the relation $ du = N(l) dl$. (Note that if we try to
  make such a transformation in four dimensions with $N=N(l, x^a)$, we shall,
  in general, obtain nonzero mixed terms $g_{1a}du\, dx^a$ in the linear
  element. Therefore, to obtain correct relations, we fix $x^a$ assuming that
  all derivatives in $x^a$ in our equations have already been calculated.)
  We thus have at fixed $x^a$
\beq
	ds_2^2 = A(u) dt^2 - du^2/A(u),
\eeq
  where $A(u) \equiv N^2(l)$. This coordinate has the following
  important properties (see, e.g, \cite{vac1,cold,bz3}): it always takes a
  finite value $u = u_h$ at a Killing horizon where $A(u) =0$;\footnote
       {In principle, $u$ can take an infinite value at a candidate horizon
        where $A\to 0$, but then, as one can check, the canonical parameter
        of the geodesics also tends to infinity, so that the space-time is
  	already geodesically complete and no continuation is required.
	Such cases (``remote horizons'') are possible in principle, but we
	will not discuss them here.	}
  moreover, near a horizon, the increment $u - u_h$ is a multiple (with a
  nonzero constant factor) of the corresponding increments of manifestly
  well-behaved Kruskal-type null coordinates, used for analytic extension
  of the metric across the horizon. Thus a horizon corresponds to a regular
  zero of $A(u)$, i.e.,
\beq                                                         \label{hor}
      A (u) \equiv N^2 (l) \sim (u-u_h)^n,
\eeq
  where $n \in \N$ is the order of the horizon. A simple, Schwarzschild-like
  horizon corresponds to $n=1$, an extremal one (as in extremal
  Reissner-Nordstr\"om \bhs) to $n=2$, and horizons with $n > 2$ are called
  ultraextremal.

  Thus all metric functions should be analytic (or smooth at least up to
  derivatives of a certain order $s\geq 2$) functions of $u$ at $u=u_h$.
  In particular, this applies to $\gamma_{ab}$ which behave as scalars in
  the $(t,u)$ subspace. Then
\beq                                                       \label{K(u)}
     K_{ab} = - \Half \frac{\d \gamma_{ab}}{\d l}
     	    = - \Half \sqrt{A} \frac{d \gamma_{ab}}{du}
	        \sim (u-u_h)^{n/2};
\eeq
  here and henceforth expressions in terms of $u$ are valid at fixed $x^a$.
  Thus $K_{ab} = O(N) \to 0$ at the horizon.

  Furthermore, assuming a Taylor expansion for $A(u)$ near $u=u_h$,
  it is easy to find the near-horizon behavior of $N(l)$:
  in accord with \cite{viss, v, tr}, we obtain
\bearr                                                     	\label{h1}
    	n = 1: \cm  l\to 0, \ \ \ \
			N(l) = \kappa_1 l + \kappa_3 l^3 + \ldots;
\yyy                                                       	\label{h2}
	n = 2: \cm  l \to \infty,\ \ \
			N(l) = B_1 \e^{-bl} + B_2 \e^{-2bl} + \ldots;
\yyy                                                            \label{h3}
	n > 2: \cm  l \to \infty,\ \ \
	     N(l) = C_n \e^{-n/(n-2)} + C_{n+2} \e^{-(n+2)/(n-2)} +\ldots,
\ear
  where the quantities $\kappa_i,\ b,\ B_i,\ C_i$ are, in general, functions
  of $x^2,\ x^3$, since these expansions have been obtained for fixed $x^a$.

  It has been shown, however \cite{v, tr}, that the quantites $\kappa_1$,
  $B_1$ and $C_n$ are constants. For simple horizons, $\kappa_1 = \kappa$ is
  nothing else but the surface gravity, and its constancy along the horizon
  is the content of the so-called zeroth law of \bh\ thermodynamics
  \cite{FroNov}. For higher-order horizons ($n \geq 2$), the constancy of
  $B_1$ and $C_n$ has been proved in \cite{v, tr} from the horizon regularity
  requirement: if one admits their $x^a$ dependence, one obtains the
  so-called truly naked horizons, at which some curvature components in the
  freely falling reference frame are infinite even though all
  algebraic curvature invariants remain finite.

  Due to $x^a$-independence of the first terms in the expansions
  (\ref{h1})--(\ref{h3}), it is possible to introduce the quasiglobal
  coordinate $u$ in our four-dimensional space-time as a whole, not only on
  its ($t, l$)-sections, at least near the horizons, and preserve the whole
  reasoning connected with regularity in terms of $u$. Thus, e.g., for simple
  horizons we have
\[
    l = \sqrt{\kappa/2} \sqrt{u-u_h} [1 + l_3(x^a)(u-u_h) + \ldots],
\]
  so that an $x^a$-dependence is absent in the main approximation with
  respect to $u-u_h$; in the same approximation, we can also neglect the
  possible $x^a$-dependence of $u_h$ (though, it does not affect our
  reasoning in any way).

\section{Matter compatible with a horizon}

  From the expressions (\ref{h1})--(\ref{h3}) it follows that $N'/N$ is
  zero at the horizon for $n=1$ and $n>2$, and it has there a finite value
  for $n=2$. From the constancy of  $\kappa,\ B_1$ and $C_1$ it also follows
  that $\Delta_2 N/N \to 0$ as $N\to 0$. Together with (\ref{K(u)}) this
  implies $G_{1}^{1}-G_{0}^{0}=0$ at the horizon, whence it follows
\bearr
	p_1 + \rho =0     				     \label{vac}
\ear
  at the horizon, in accord with \cite{viss} (though our proof is different
  from that in \cite{viss}).

  Let us assume that (at least near the horizon) the equation of state of
  matter takes the form
\beq
	p_1\approx w\rho  					\label{w}
\eeq
  with $w = \const$. We also suppose that
\beq                                                         \label{T_ab}
	|T_{a}^{b}|/\rho < \infty.
\eeq

  In the conservation law (\ref{cons}), the ratio of the third term to the
  second one is of the order $(\max K_{ab}) N/N'$. For simple horizons ($n=1$)
  $N'> 0$, and in the extremal case ($n=2$) $N'\sim N$ whereas $K_{ab} \sim N$
  (at most). Thus in both these cases we can safely neglect the third term
  in (\ref{cons}). In the ultraextremal case ($n>2$) $N/N' \sim l \to \infty$;
  however, due to (\ref{h3}) the ratio in question behaves as $l^{-2/(n-2)}
  \to 0$ as $l\to \infty$, so that in this case the third term in
  (\ref{cons}) is also negligible as compared with the second one.

  With the third term neglected, \eqs (\ref{cons}) and (\ref{w}) lead to
\beq
	\rho \sim N^{-(1+w)/w}.  			     \label{rho}
\eeq
  Thus the admissible range of $w$, in which $\rho <\infty$ at the horizon,
  is $-1 \leq w < 0$, where $w=-1$ corresponds to the so-called vacuum fluid
  \cite{dym92} whose definitive property is that in a certain spatial
  direction $x^1$ there holds the relation (\ref{vac}). It generalizes the
  notion of a \sph\ vacuum fluid (see also \cite{bz2} and references
  therein) and certainly also includes the cosmological constant as a special
  case.

  For other values of $w$, we have $\rho \to 0$ as $N \to 0$.

  On the other hand, as before, we can consider the Einstein equations
  (\ref{1-0}) and (\ref{11}) at fixed values of the coordinates $x^a$,
  and then, since all metric functions should depend analytically (or
  sufficiently smoothly) on the quasiglobal coordinate $u$, the same is true
  for the right-hand side of these equations, which, for $w > -1$, means that
  $\rho \sim (u-u_h)^k$ where $k$ is a positive integer. Comparing this
  condition with (\ref{hor}) and (\ref{rho}), we obtain
\beq                                                         \label{w_nk}
       w = -\frac{n}{n+2k},
\eeq
  thus reproducing the result of \cite{bz2}, obtained there for \sph\
  configurations. Thus matter which can be in equilibrium with static
  horizons of any shape should have an equation of state with the parameter
  $w$ from the discrete set of values (\ref{w_nk}), where $n$ is the order of
  the horizon and $k$ characterizes the near-horizon behavior of the density.

  We can also admit a non-interacting mixture of our matter and a vacuum
  fluid with the property $\rho\vac + p_1\vac = 0$, The conservation law
  (\ref{cons}) then holds for each of them separately, and nothing is changed
  in the above reasoning.

  One more restriction follows from \eq (\ref{11}). Indeed,
  at the horizon itself it gives
\beq                                                          \label{hh}
	\frac{1}{2} R_{\parallel } + K_1 = 8\pi \rho\vac,\cm
		K_1 := \lim_{\rm hor}\frac{KN'}{N}.
\eeq
  Generically (if $K$ does not vanish quicker than $N$ when approaching the
  horizon), we have $K_1 \sim (u-u_h)^{n-1}$.

  Further conclusions depend on the presence of a vacuum fluid.

\medskip\noi
{\bf 1.} $\rho\vac = 0$.
  In the general case that the horizon is curved, $R_{\parallel} \ne 0$,
  according to (\ref{hh}), we must have $K_{1}\neq 0$, so that the horizon
  must be simple, $n=1$. Higher-order horizons ($n \geq 2$) are only possible
  if $R_{\parallel }=0$. In particular, they cannot exist in \sph\
  space-times, in accord with \cite{bz2}, but they can appear in
  cylindrically symmetric black holes since the corresponding two-dimensional
  cross-section is flat for them.

\medskip\noi
{\bf 2.}  $\rho_{vac}\neq 0$. \eq (\ref{hh}) admits any order of the horizon
  but, if it is higher-order, $K_1 =0$, hence flat horizons ($R_{\parallel}
  =0$) are now impossible. For $n \geq 2$, \eq (\ref{hh}) gives a direct
  relationship between the horizon curvature and the vacuum fluid density
\[
	R_{\parallel} = 16\pi \rho\vac.
\]

  Thus, without knowing any exact solutions, we have obtained certain
  restrictions on the possible horizon geometry depending on the existence
  and properties of the vacuum fluid.

\section{Concluding remarks}

  We have shown that the main result of \cite{bz2}, obtained there for
  Killing horizons in \ssph\ space-times, also holds for horizons in general
  static space-times: an equilibrium between a horizon and matter is only
  possible for some particular values of the parameter $w = p_1/\rho$ given
  by \eq (\ref{w_nk}). In the generic situation of a simple (non-extremal)
  horizon and the slowest possible density decrease near the horizon, this
  corresponds to $w = -1/3$, the value characteristic of a gas of disordered
  cosmic strings.

  A new feature as compared to spherical symmetry is that, in the absence of
  vacuum matter, higher-order horizons are possible provided the
  horizon as a 2-surface has zero scalar curvature $R_{\parallel}$. In
  particular, this is true for cylindrical black holes. If $\rho\vac\neq
  0$, the sign of $\rho\vac$ should coincide with that of $R_{\parallel}$.
  In particular, $\rho_{\rm vac} < 0$, forbidden in the spherical case, is
  compatible with a horizon with hyperbolic geometry. Thus matter with
  certain negative values of $w$ can contain horizons with non-spherical
  topologies (cf. \cite{un}).

  In our reasoning, which has been entirely local and relied on near-horizon
  expansions, we did not assume any particular equation of state for matter
  and even did not restrict the behaviour of the transverse pressure except
  for its regularity requirement. Moreover, no assumptions on the asymptotic
  properties of the space-time manifold were made.  In this sense, our
  conclusions are model-independent.

\end{document}